\documentclass[%
 reprint,
 amsmath,amssymb,
 aps,
prc,
]{revtex4-2}

\usepackage{graphicx}
\usepackage{dcolumn}
\usepackage{bm}
\usepackage{hyperref}
\usepackage{tikz}
\usetikzlibrary{positioning,arrows.meta}
\usepackage{helvet}
\usepackage{sansmath}

\begin{document}

\title{\texorpdfstring{Constraining the Low-$\bm{p_T}$ $\bm{\eta/\pi^0}$ Ratio for Direct-Photon Analyses with Blast-Wave Fits to $\bm{\pi}$, $\bm{K}$, and $\bm{p}$ Spectra}{Constraining the Low-p\_T eta/pi0 Ratio for Direct-Photon Analyses with Blast-Wave Fits to pi, K, and p Spectra}}

\author{Klaus Reygers}%
 \email{reygers@physi.uni-heidelberg.de}
\affiliation{Physikalisches Institut, Universit\"{a}t Heidelberg, 69120 Heidelberg, Germany}
\author{Andreas Kirchner}
\email{andreas.kirchner@duke.edu}
\affiliation{Department of Physics, Duke University, Durham, North Carolina 27708, USA}
\affiliation{Institut f\"{u}r Theoretische Physik, Universit\"{a}t Heidelberg, 69120 Heidelberg, Germany}
\author{Aleksas Mazeliauskas}%
 \email{a.mazeliauskas@thpys.uni-heidelberg.de}
\affiliation{Institut f\"{u}r Theoretische Physik, Universit\"{a}t Heidelberg, 69120 Heidelberg, Germany}



\date{September 27, 2026}

\begin{abstract}
We predict the $\eta/\pi^0$ ratio at low $p_T$ ($p_T \lesssim 3~\mathrm{GeV}/c$) using the measured charged $K/\pi$ ratio and model input from a blast-wave framework with feeddown contributions. This approach can provide improved, data-constrained background estimates for direct-photon and dilepton measurements in heavy-ion collisions. In this approach, the explicit modeling of radial flow and hadronic feeddown enables an uncertainty estimate for the low-$p_T$ extrapolation of the $\eta/\pi^0$ ratio. Using central Pb--Pb collisions at $\sqrt{s_{NN}}=2.76$~TeV as an example, we find that the $\eta$-related decay-photon uncertainty at $p_T \approx 1~\mathrm{GeV}/c$ is of order 10\% of the expected direct-photon signal.
\end{abstract}

\maketitle

\section{Introduction}
Ultrarelativistic collisions of nuclei are a unique laboratory for studying quantum chromodynamics (QCD) at extreme temperatures and densities. These collisions are widely interpreted as creating a quark-gluon plasma (QGP), i.e., an approximately thermalized medium whose relevant degrees of freedom are quarks and gluons \cite{Busza:2018rrf}. The QGP expands, cools, and hadronizes at a temperature of about 156~MeV~\cite{Borsanyi:2018grb,HotQCD:2018pds}, which coincides with the experimentally extracted chemical freeze-out temperature~\cite{Andronic:2017pug}. Real and virtual photons (dileptons) are a key probe of the medium as they are produced in all stages of the collision and are not affected by hadronization. Thermal photons and dileptons are of particular interest since their spectrum reflects the temperature of the medium. 

Experimentally, one measures direct photons by statistically subtracting the large background of photons from hadron decays. The largest contribution to the decay-photon background comes from the $\pi^0 \to \gamma \gamma$ decay (about 83--88\%), followed by the $\eta \to \gamma \gamma$ decay (about 10--15\%). Together, the $\pi^0$ and the $\eta$ contribute about 98\% to the decay-photon background. The third-largest contribution, up to about 1--2\%, comes from the decays $\omega \to \pi^0 \gamma (\eta \gamma)$. All other hadron decays are practically negligible~\cite{ALICE:2015xmh}. In the case of dileptons, a similar experimental approach of subtracting a ``cocktail'' of known background sources from the total number of correlated $e^+e^-$ pairs is taken. Here the Dalitz decays $\pi^0 \to \gamma e^+e^-$ and $\eta \to \gamma e^+e^-$ are the most important background sources at low $e^+e^-$ invariant masses.

For real photons, the low transverse momentum region ($p_T \lesssim 3\,\mathrm{GeV}/c$) is of special interest due to the significant contribution of thermal photons from the QGP phase and the subsequent hadron gas phase~\cite{David:2019wpt}. The yield of background photons from the two-photon decay of the $\eta$ meson is expected to be similar in magnitude or larger than the signal of thermal photons. However, at low $p_T$ the measurement of the $\eta$ meson is difficult and typically associated with large uncertainties. Therefore, various approaches to model the $p_T$ spectrum of the $\eta$ meson were employed in the past. No approach yet combines experimental constraints with decay feeddown in a unified framework.

In this paper, we present an approach to model the $\eta/\pi^0$ ratio at low $p_T$ using measured charged $K/\pi$ input and blast-wave modeling of hadron spectra, including decay contributions from short-lived hadrons \cite{Mazeliauskas:2019ifr}. Our central claim is that the new method enables improved control and reduction of the decay-cocktail uncertainties associated with the $\eta$ meson. This is expected to enable a more precise extraction of the direct-photon and low-mass dilepton signals in heavy-ion collisions.

\section{\boldmath Previous Approaches to Constraining Low-$p_T$ $\eta/\pi^0$}

Different approaches to model the $\eta/\pi^0$ ratio at low $p_T$ were used in previous measurements of direct photons in nucleus--nucleus collisions. In Pb--Pb collisions at $\sqrt{s_\mathrm{NN}} = 17.3\,\mathrm{GeV}$ at the CERN SPS, the WA98 experiment assumed $m_T$ scaling to estimate the $\eta/\pi^0$ ratio in the unmeasured low-$p_T$ region \cite{WA98:2000vxl}. The $m_T$ scaling hypothesis states that invariant hadron yields for different particle species have the same spectral shape as a function of $m_T = \sqrt{m^2+p_T^2}$ and only differ by an overall normalization factor \cite{Albrecht:1991zza}. The $m_T$ scaling hypothesis was also used in direct-photon measurements by the PHENIX experiment at RHIC \cite{PHENIX:2008uif}. At high collision energies, however, the strong collective radial flow of the medium is expected to give rise to significant deviations from $m_T$ scaling. At the LHC, the ALICE experiment considered the $m_T$ scaling prediction as a limiting case to estimate the systematic uncertainty of the $\eta/\pi^0$ ratio \cite{ALICE:2015xmh}. 

To take the effect of radial flow into account, the STAR experiment used a Tsallis blast-wave model to predict the $\eta/\pi^0$ ratio \cite{STAR:2016use}. More recently, Ren and Drees proposed a method to estimate the $\eta/\pi^0$ ratio in nucleus--nucleus (AA) collisions based on the assumption that $\eta/\pi^0$ as a function of $p_T$ is approximately independent of the collision energy over a wide range of energies \cite{Ren:2021pzi}. The effect of radial flow on the $\eta/\pi^0$ ratio is then modeled by multiplying the universal $\eta/\pi^0$ in proton--proton collisions by the ratio of the charged $K/\pi$ ratios in AA and pp collisions:
\begin{equation}
\left(\eta/\pi^0\right)_{\mathrm{AA}} =
\frac{\left(K/\pi\right)_{\mathrm{AA}}}{\left(K/\pi\right)_{\mathrm{pp}}}
\left(\eta/\pi^0\right)_{\mathrm{pp}}.
\end{equation}
This approach was used in a recent direct-photon measurement by the PHENIX experiment \cite{PHENIX:2022rsx}. 

Both the Tsallis blast-wave approach by the STAR experiment and the Ren--Drees method underline the importance of modeling the effect of radial flow for the prediction of the $\eta/\pi^0$ ratio. A blast-wave model approach, as used by STAR, can be improved by including the effect of decays of short-lived hadrons. This applies also to the method by Ren and Drees where kaons are used as a flow proxy for $\eta$. This implicitly assumes that the effect of resonance decays is similar for the $\eta/\pi^0$ ratio and the $K/\pi$ ratio. Moreover, the effect of radial flow depends on the mass of the particle. The mass difference between the kaon ($m_K = 494\,\mathrm{MeV}$) and the $\eta$ meson ($m_\eta = 548\,\mathrm{MeV}$) could therefore affect the accuracy of this method. We will compare the results of the Ren--Drees approach and our approach in this paper.

\section{\boldmath Method: Data-Driven $\eta/\pi^{0}$ from Measured $K/\pi$}

We determine the low-$p_T$ ($p_T \lesssim 3\,\mathrm{GeV}/c$) $\eta/\pi^0$ ratio in nucleus--nucleus collisions using the measured $K/\pi$ ratio in the same centrality class. This strategy exploits the fact that, unlike $\eta/\pi^0$, $K/\pi$ can be measured directly with high precision at low $p_T$. Our core method models hadron spectra within a blast-wave framework that includes resonance feeddown. In addition, we consider flow scaling as an alternative way to predict hadron spectra, which can be viewed as a generalization of $m_T$ scaling in the presence of strong radial flow. To our knowledge, this is the first application of this flow-scaling construction to model hadron spectra in heavy-ion collisions.

\subsection{Core Double-Ratio Method}
The central idea of our core double-ratio method is to determine the $\eta/\pi^0$ ratio by multiplying the model ratio $[(\eta/\pi^0)/(K/\pi)]_\mathrm{model}$ by the measured charged $K/\pi$ ratio. By constructing $\eta/\pi^0$ based on $K/\pi$, common normalization uncertainties cancel. The model parameters are determined by a fit to the measured $\pi$, $K$, and $p$ spectra. To keep this task manageable, we employ a blast-wave model \cite{Schnedermann:1992ra,Schnedermann:1993ws} based on the Cooper-Frye freeze-out prescription \cite{Cooper:1974mv}:
\begin{equation}
\left(\eta/\pi^0\right)_{\mathrm{AA}}
=
\frac{\left(\eta/\pi^0\right)_{\mathrm{blast\text{-}wave}}}
     {\left(K/\pi\right)_{\mathrm{blast\text{-}wave}}}
\left(K/\pi\right)_{\mathrm{AA,\,meas}} \, .
\label{eq:eta_to_pi0_ratio_from_data_and_blastwave}
\end{equation}
The free parameters of the blast-wave model are the kinetic freeze-out temperature $T_\mathrm{kin}$, the surface velocity $\beta_s$, and the radial flow profile exponent $n$. We calculate the blast-wave spectra with Bose/Fermi quantum statistics using a truncated series expansion. This mainly matters at low transverse momentum, in particular for pions. Concretely, for species $a$ we use
\begin{align}
\frac{dN_a}{dp_T}
&\propto
p_T \sum_{k=1}^{N_q} s_a^{\,k-1}
\int_0^1 d\hat r\;\hat r\, m_T\,
I_0\!\left(k\,\frac{p_T\sinh\rho(\hat r)}{T_\mathrm{kin}}\right)
\notag\\[-0.2em]
&\qquad\times
K_1\!\left(k\,\frac{m_T\cosh\rho(\hat r)}{T_\mathrm{kin}}\right),
\\
m_T&=\sqrt{m_a^2+p_T^2},\;
\rho(\hat r)=\mathrm{arctanh}\!\left(\beta_s \hat r^{\,n}\right),\;
\hat r\equiv r/R .
\end{align}
Here $I_0$ and $K_1$ are modified Bessel functions of the first and second kind, respectively. Furthermore, $s_a=+1$ for bosons and $s_a=-1$ for fermions, $N_q$ is the truncation order, and $g_a=2J_a+1$ is the spin degeneracy. The Boltzmann approximation corresponds to $N_q=1$.

Our blast-wave model includes feeddown from short-lived resonances such as $\rho \rightarrow \pi\pi$ and $\Delta \rightarrow N\pi$. The fit is accelerated by using precomputed feeddown kernels, similar in spirit to the FastReso approach~\cite{Mazeliauskas:2018irt,FastReso}. For a decay chain from parent species $a$ to daughter species $b$, the Lorentz-invariant spectra are related by
\begin{equation}
E_{\bm p}\frac{dN_b}{d^3p}
=
\int \frac{d^3q}{(2\pi)^3\,2E_{\bm q}}\,
D_b^a(\bm{p},\bm{q})\,
E_{\bm q}\frac{dN_a}{d^3q}\,.
\label{eq:decay_map_invariant}
\end{equation}
For azimuthally symmetric spectra, this is reduced to a one-dimensional transverse-momentum kernel form,
\begin{equation}
\frac{dN_b}{dp_T}
=
\int dq_T\,
A_b^a(p_T,q_T)\,
\frac{dN_a}{dq_T}\,.
\label{eq:decay_map_pt}
\end{equation}
In practice, we discretize Eq.~\eqref{eq:decay_map_pt} as a matrix multiplication over parent and daughter $p_T$ bins. The kernel element $A_{ij}$ gives the daughter yield in daughter-$p_T$ bin $i$ per parent produced in parent-$p_T$ bin $j$, with normalization performed separately in each parent-$p_T$ bin. The kernels are constructed from PYTHIA~8 decay simulations~\cite{Sjostrand:2014zea}. We include hadrons up to $m=2\,\mathrm{GeV}/c^2$ as primary parent candidates and treat particles with proper decay length $c\tau < 1\,\mathrm{mm}$ as unstable.

For this work, we augment the default PYTHIA hadron content with additional hadron states from the Thermal-FIST~\cite{Vovchenko:2019pjl} PDG-2025 hadron list that are not present in the default PYTHIA particle table and satisfy the same selection criteria, $m<2\,\mathrm{GeV}/c^2$ and $c\tau<1\,\mathrm{mm}$. These states are implemented in PYTHIA through an external particle-definition and decay-channel overlay. For $\eta$ feeddown, the most relevant added parent states are $\eta(1295)$, $N(1535)$, and $\eta(1405)$. The relative normalization of primary yields is obtained from the Grand-Canonical statistical-model particle density of species $a$ with mass $m_a$ and spin degeneracy $g_a$, given by~\cite{Cleymans:1999st}
\begin{equation}
n_a = \frac{g_a}{2\pi^2}\,T_\mathrm{ch}\,m_a^2
\sum_{k=1}^{n}
\frac{s_a^{\,k-1}}{k}\,
K_2\!\left(\frac{k\,m_a}{T_\mathrm{ch}}\right)\,
e^{\,k\mu_a/T_\mathrm{ch}} \,.
\end{equation}
Here $s_a$ is the Bose/Fermi sign defined above, and $K_2$ is the modified Bessel function of the second kind of order 2. For heavy-ion collisions at the LHC we use $T_\mathrm{ch} = 156~\mathrm{MeV}$ and a chemical potential $\mu_a = 0$~\cite{Andronic:2017pug}.

\subsection{\boldmath Parameterization of the $\eta/\pi^0$ ratio at low and high transverse momentum}

\begin{figure}[t]
\centering
\begingroup
\sansmath
\resizebox{\columnwidth}{!}{%
\begin{tikzpicture}[
  node distance=9mm and 3mm,
  box/.style={draw, align=center, inner sep=3pt, font=\sffamily\small},
  arr/.style={-Triangle, line width=0.8pt}
]
\node[box] (data) {Measured spectra\\$\pi^\pm, K^\pm, p+\bar p$};
\node[box, right=of data] (fit) {Blast-wave fit\\($T_{\mathrm{kin}}, \beta_s, n$)};
\node[box, right=of fit] (kernel) {Feeddown kernels\\(PYTHIA decay chains)};
\node[box, below=of fit] (dr) {Model double ratio\\$\left[(\eta/\pi^0)/(K/\pi)\right]_{\mathrm{model}}$};
\node[box, left=of dr] (kpi) {Measured $K/\pi$};
\node[box, right=of dr] (eta) {Pseudo low-$p_T$\\$\eta/\pi^0$ points};
\node[box, below=of dr] (final) {Final $\eta/\pi^0$ fit\\(data + pseudo)};
\node[box, below=of final] (cocktail) {Improved decay-photon cocktail};

\draw[arr] (data) -- (fit);
\draw[arr] (kernel) -- (fit);
\draw[arr] (fit) -- (dr);
\draw[arr] (kpi) -- (dr);
\draw[arr] (dr) -- (eta);
\draw[arr] (eta) -- (final);
\draw[arr] (final) -- (cocktail);
\end{tikzpicture}
}
\endgroup
\caption{Workflow for constructing a data-constrained low-$p_T$ $\eta/\pi^0$ ratio and propagating it to the decay-photon cocktail in the blast-wave model + feeddown approach.}
\label{fig:workflow}
\end{figure}
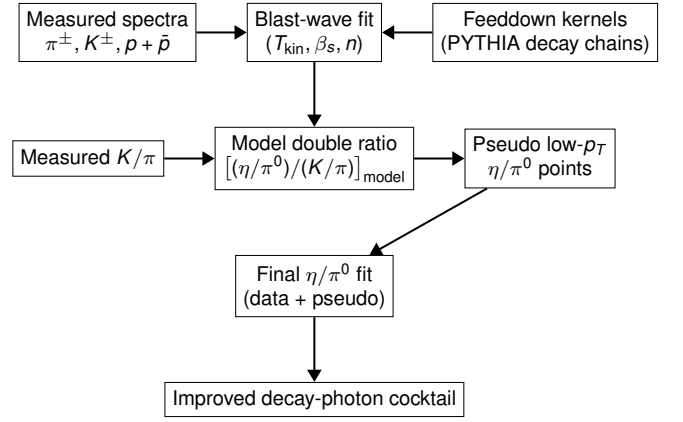

Eq.~\ref{eq:eta_to_pi0_ratio_from_data_and_blastwave} provides pseudo-data points for $p_T \lesssim 3~\mathrm{GeV}/c$ which are then used along with the measured $\eta/\pi^0$ data points to obtain a parameterization over a large transverse momentum range. For $p_T \gtrsim 5\,\mathrm{GeV}/c$ the $\eta/\pi^0$ ratio reaches an approximately constant value of $r = 0.487 \pm 0.024$ in nucleus--nucleus collisions, independent of collision energy and centrality \cite{Ren:2021pzi}. Based on the simple parameterization of a thermal source with temperature $T$ and radial flow velocity $\beta$
\begin{equation}
  \frac{dN}{dp_T\,dy} \propto \exp\left(\frac{\gamma (\beta p_T - m_T)}{T} \right) \equiv f_{m, \beta, T}(p_T)
\end{equation}
where the transverse mass $m_T = \sqrt{m^2 + p_T^2}$ and $\gamma = (1 - \beta^2)^{-1/2}$, 
we parameterize the $\eta/\pi^0$ ratio as
\begin{equation}
 \frac{\eta}{\pi^0}(p_T) = s(p_T) A \frac{f_{m_\eta, \beta, T}(p_T)}{f_{m_{\pi^0}, \beta, T}(p_T)} + (1 - s(p_T)) r \,.
 \label{eq:eta_to_pi0_parameterization}
\end{equation}
where $A$ is a normalization parameter. We treat $r$ as a fixed parameter. The function
\begin{equation}
  s(p_T) = 1 - \frac{1}{2} \left(1 + \mathrm{erf}\left(\frac{p_T - p_0}{d}\right)\right) 
\end{equation}
provides a smooth transition from a region where the hydrodynamic description applies to the $p_T$-independent value $r$ at high $p_T$. The parameters $p_0$ and $d$ define the location and smoothness of the transition, respectively.
The flow-inspired parameterization of Eq.~\ref{eq:eta_to_pi0_parameterization} is flexible enough to describe a possible maximum in the $\eta/\pi^0$ ratio before it approaches a constant asymptotic value, similar to the maximum observed for the $K/\pi$ ratio in central Pb--Pb collisions at $\sqrt{s_{NN}}=2.76$~TeV \cite{ALICE:2014juv}. At the same time, the approximately $\sqrt{s}$-independent $\eta/\pi^0$ ratio in pp collisions can be well described with $\beta = 0$. The entire workflow of our approach is summarized in Fig.~\ref{fig:workflow}.

\subsection{Flow-Scaling Alternative}
We introduce the flow-scaling method as an alternative cross-check to our core double-ratio method. Using a single effective flow velocity $\beta$, we assume that hadron $p_T$ spectra for species $a$ can be approximated by
\begin{equation}
  \frac{dN_a}{dp_T\,dy} = C_a F(m_T^*)
\end{equation}
where $F$ is a universal scaling function describing the shape and the Lorentz-transformed transverse mass $m_T^*$ is defined as
\begin{equation}
  m_T^* = \gamma (m_T - \beta p_T) \,.
\end{equation}
Based on the measured $\pi$, $K$, $p$ spectra, we determine the optimal flow velocity $\beta$ in the range $0.5 < p_T < 2.5\,\mathrm{GeV}/c$ where spectra are well described by hydrodynamic models.

In the flow-scaling method we can predict the $\eta/K$ ratio as a function of $p_T$ as 
\begin{equation}
  (\eta/K)_\mathrm{flow\text{-}scaling}
  = 
  \frac{F\left(\gamma (\sqrt{m_\eta^2+p_T^2} - \beta p_T)\right)}{F\left(\gamma (\sqrt{m_K^2+p_T^2} - \beta p_T)\right)}, 
\end{equation}
i.e., in this work we assume $C_\eta = C_K$. To compare this result with the $\eta/\pi^0$ ratio as a function of $p_T$ from the double-ratio method, we use the $\pi/\pi^0$ ratio from the double-ratio method according to
\begin{equation}
  \begin{split}
  (\eta/\pi^0)_\mathrm{flow\text{-}scaling}
  &= (\eta/K)_\mathrm{flow\text{-}scaling}\,(\pi/\pi^0)_\text{blast-wave} \\
  &\quad\times \left(K/\pi\right)_{\mathrm{AA,\,meas}} \, .
  \end{split}
  \label{eq:eta_to_pi0_from_eta_to_K}
\end{equation}


\section{\boldmath Inputs and Model Ingredients (Pb--Pb $\sqrt{s_{NN}}=2.76$ TeV, 0--10\%)}

\subsection{\boldmath Simultaneous Fit to $\pi$, $K$, and $p$ Spectra}
\begin{figure}[t]
\centering
\includegraphics[width=\columnwidth]{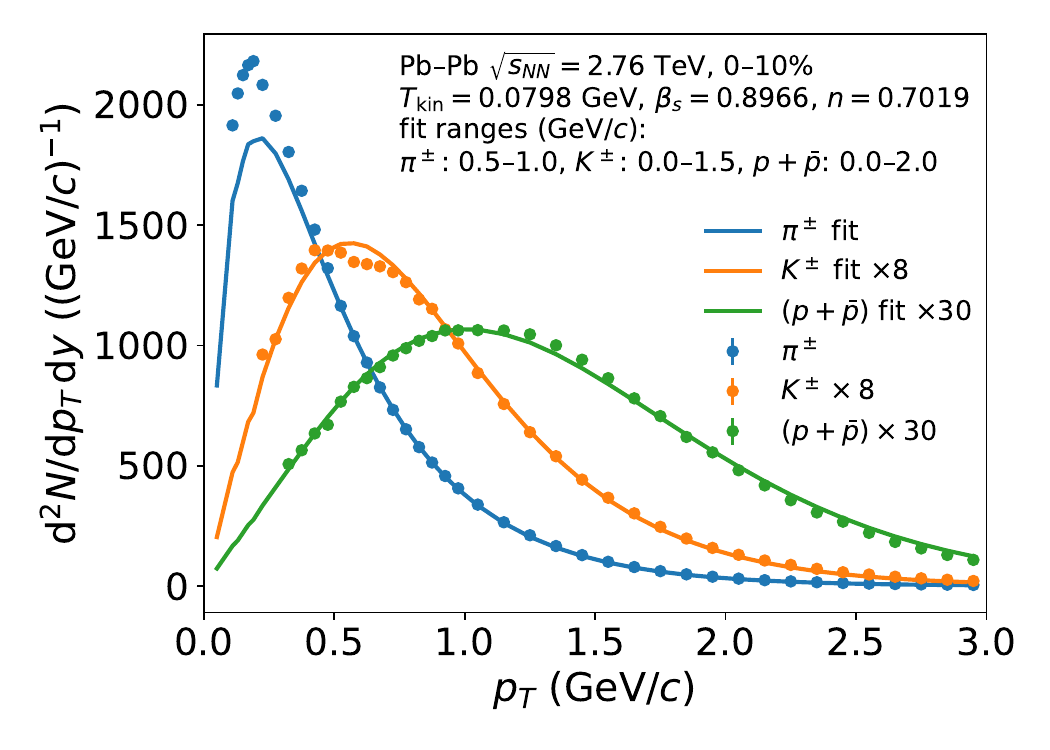}
\caption{Simultaneous blast-wave + feeddown fit to the measured $\pi^\pm$, $K^\pm$, and $(p+\bar p)$ spectra in 0--10\% Pb--Pb collisions at $\sqrt{s_{NN}}=2.76$ TeV \cite{ALICE:2014juv}. For visibility, the $K^\pm$ and $(p+\bar p)$ spectra are scaled by factors of 8 and 30, respectively.}
\label{fig:fit_blastwave_2p76_0_10}
\end{figure}

We apply our method to $\pi$, $K$, and $p$ spectra measured by ALICE at $\sqrt{s_{NN}}=2.76$ TeV in the 0--10\% centrality class. The simultaneous blast-wave fit including feeddown is shown in Fig.~\ref{fig:fit_blastwave_2p76_0_10}. The relative normalizations of the $\pi$, $K$, and $p$ spectra are left as free parameters. Even though feeddown is included, the model underestimates the pion spectrum at low $p_T$. This has been observed to be a rather generic feature when describing pion spectra with hydrodynamic models~\cite{Lu:2024shm,Nijs:2020roc}. In the present analysis, however, we are primarily concerned with $p_T \gtrsim 0.5\,\mathrm{GeV}/c$, so this low-$p_T$ mismatch is not a limiting issue for our strategy.

\subsection{Double Ratio from Blast-Wave Model with Feeddown}

\begin{figure}[t]
\centering
\includegraphics[width=\columnwidth]{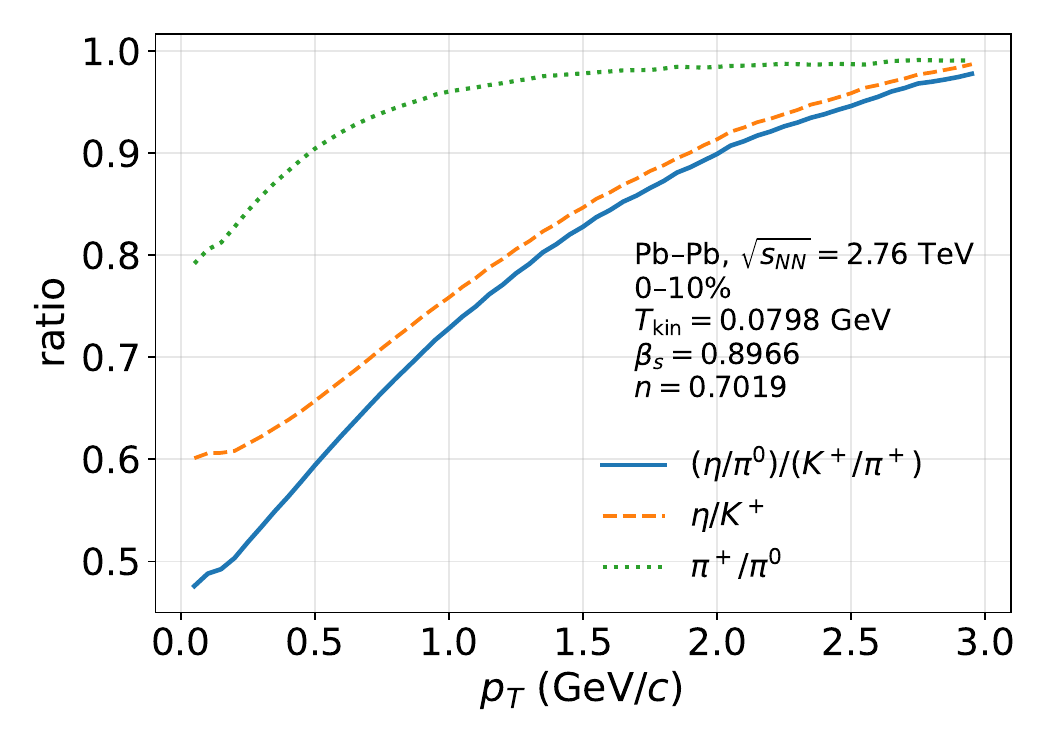}
\caption{Feeddown-based blast-wave model double ratio $(\eta/\pi^0)/(K^+/\pi^+)$ for 0--10\% Pb--Pb collisions at $\sqrt{s_{NN}}=2.76$ TeV (solid line). The component ratios $\eta/K^+$ (dashed) and $\pi^+/\pi^0$ (dotted) are shown for reference.}
\label{fig:model_double_ratio_2p76_0_10}
\end{figure}

The double ratio $(\eta/\pi^0)/(K^+/\pi^+)$ for the blast-wave parameters obtained from the fit to the $\pi$, $K$, and $p$ spectra is shown in Fig.~\ref{fig:model_double_ratio_2p76_0_10}. We also show the $\eta/K^+$ ratio and the $\pi^+/\pi^0$ separately. The larger $\pi^0$ yield at low $p_T$ compared to $\pi^+$ is related to isospin-violating decays of the $\eta$ meson.

Propagating the blast-wave fit covariance yields a double-ratio uncertainty band smaller than the plotted line width and it is therefore not shown in Fig.~\ref{fig:model_double_ratio_2p76_0_10}.

\section{Cross-Checks and Systematic Uncertainties}
\label{sec:systematics}
We estimate the systematic uncertainty of the predicted $(\eta/\pi^0)/(K^+/\pi^+)$ by variations performed within the blast-wave + feeddown framework. Specifically, we vary the fit ranges of the measured pion, kaon, and proton spectra, repeat the fit using total rather than statistical experimental uncertainties, and change the chemical freeze-out temperature in the statistical-model particle densities from the baseline value of 156~MeV to 136 and 176~MeV. The spectra are refitted for every variation, so changes in the inferred radial-flow and freeze-out parameters and in the feeddown composition are propagated consistently to the double ratio.

\begin{figure}[t]
\centering
\includegraphics[width=\columnwidth]{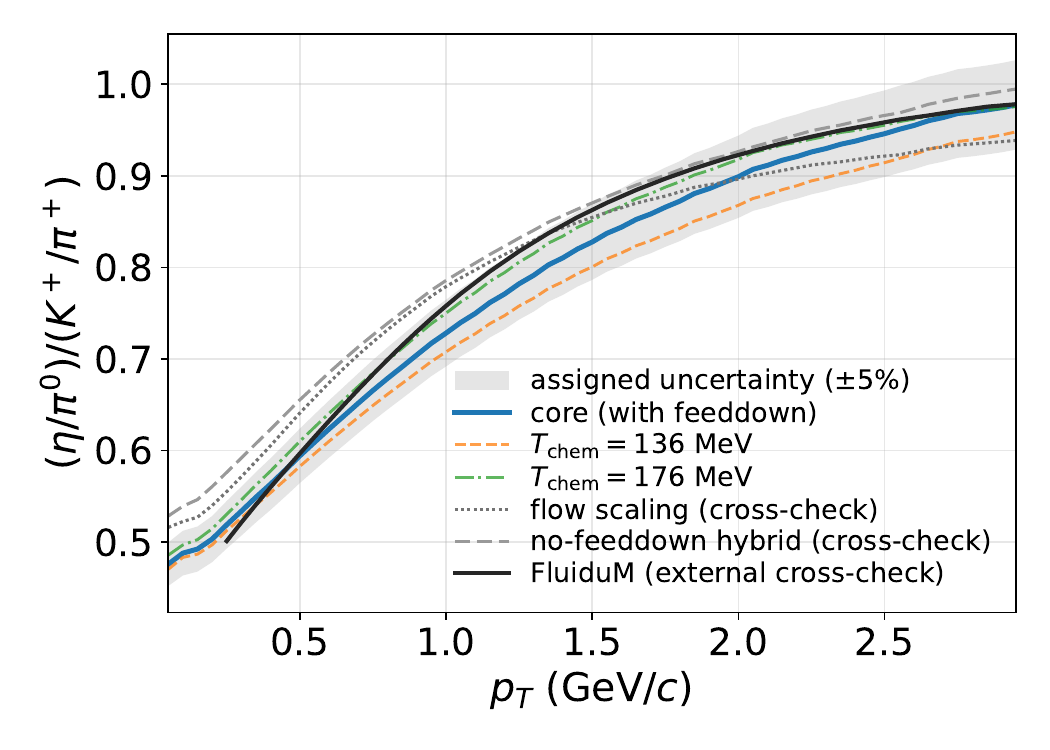}
\caption{Systematic study of the double ratio $(\eta/\pi^0)/(K^+/\pi^+)$ in 0--10\% Pb--Pb collisions at $\sqrt{s_{NN}}=2.76$ TeV. The baseline blast-wave + feeddown result is shown together with both $T_\mathrm{chem}$ variations and the assigned constant uncertainty (gray band). The flow-scaling and no-feeddown-hybrid curves are qualitative cross-checks only and do not enter the uncertainty envelope. FluiduM is included as an external cross-check and likewise does not enter the envelope.}
\label{fig:systematics_summary_2p76_0_10}
\end{figure}

At each $p_T$, we form an envelope from these internal variations. In the interval $0.5 \le p_T \le 2.5~\mathrm{GeV}/c$, the largest relative deviation from the baseline is 3.5\%, dominated by the $T_\mathrm{chem}=136$~MeV variation. Given the approximate nature of this procedure, we conservatively assign a constant relative uncertainty of $\pm 5\%$ to the double ratio rather than retaining the detailed $p_T$ dependence of the envelope. The gray band in Fig.~\ref{fig:systematics_summary_2p76_0_10} shows this assigned uncertainty. Of the internal variations, only the two $T_\mathrm{chem}$ variations are shown in Fig.~\ref{fig:systematics_summary_2p76_0_10}; all internal variations enter the numerical envelope. The flow-scaling and no-feeddown-hybrid curves are also shown as qualitative cross-checks, but do not enter the systematic uncertainty because they do not implement the complete baseline feeddown treatment. A fixed linear radial-velocity profile ($n=1$) is retained as an additional stress test but is not included in the uncertainty estimate because it gives a substantially poorer description of the identified-particle spectra.

As an external cross-check, we compare the result with a full hydrodynamic simulation using FluiduM \cite{Floerchinger:2018pje}, with separate chemical and kinetic freeze-out in the partial-chemical-equilibrium approach and FastReso decay feeddown \cite{Mazeliauskas:2018irt}. The parameters for 0--10\% Pb--Pb collisions at $\sqrt{s_{NN}}=2.76$ TeV are taken from Ref.~\cite{Vermunt:2023zsk}. The FluiduM result is not used to construct the uncertainty envelope. It lies within the assigned band over the relevant interval, with a maximum deviation of 4.6\%. This agreement supports the plausibility of the internal estimate for this collision system, but does not demonstrate that the band encompasses the full dependence on the hydrodynamic evolution and freeze-out prescription. Quantifying that dependence would require a systematic comparison among several hydrodynamic models and is beyond the scope of this work. The assigned uncertainty should therefore be understood as an estimate of systematic variations within the present framework and should be re-evaluated for other collision systems or centralities.

\section{\boldmath Results: $\eta/\pi^{0}(p_T)$ in 0--10\% Pb--Pb at 2.76 TeV}

\begin{figure}[t]
\centering
\includegraphics[width=\columnwidth]{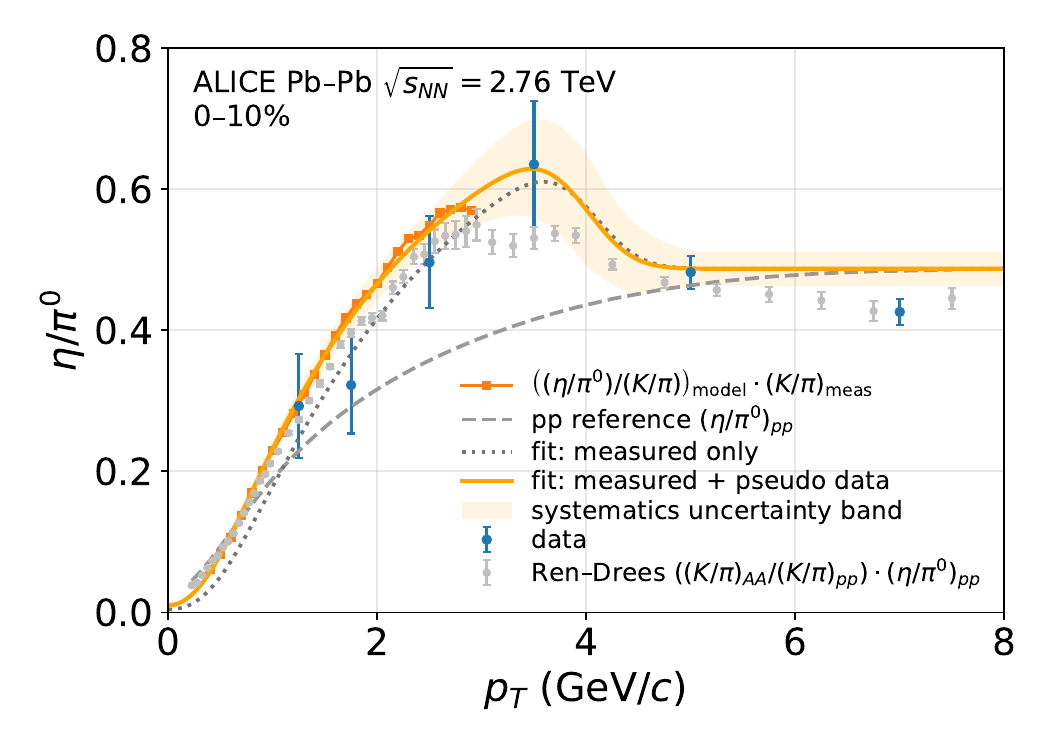}
\caption{$\eta/\pi^0$ in 0--10\% Pb--Pb collisions at $\sqrt{s_{NN}}=2.76$ TeV \cite{ALICE:2018mdl}. Measured points are combined with pseudo-data constraints from the double-ratio method and fitted with the flow-inspired parameterization of Eq.~\ref{eq:eta_to_pi0_parameterization}. The shaded band indicates the systematic uncertainty model used for the low-$p_T$ extrapolation. We also show a fit of only the measured $\eta/\pi^0$ points without the low-$p_T$ pseudo-data constraints (dark gray dashed line). The parameterization of the $\eta/\pi^0$ ratio in pp collisions is shown for reference (light gray dashed line) and illustrates the significant effect of radial flow on the $\eta/\pi^0$ ratio in heavy-ion collisions. The blast-wave + feeddown prediction shows good agreement with the Ren--Drees prediction (gray points).}
\label{fig:eta_to_pi0_2p76_0_10}
\end{figure}

We can now predict the $\eta/\pi^0$ ratio for $p_T \lesssim 3~\mathrm{GeV}/c$ by multiplying the blast-wave double ratio with the measured charged $K/\pi$ ratio. The resulting pseudo-data points are then fitted together with the measured $\eta/\pi^0$ data to obtain a parameterization over the full $p_T$ range. The resulting curve and the associated systematic uncertainty band for the low-$p_T$ extrapolation are shown in Fig.~\ref{fig:eta_to_pi0_2p76_0_10}. We compare our result to the Ren--Drees method and find that both approaches give very similar predictions for the low-$p_T$ $\eta/\pi^0$ ratio. In Sec.~\ref{sec:feeddown_structure}, we show that the feeddown contribution to both the $\eta$ spectrum (Fig.~\ref{fig:feeddown_fractions_eta}) and the $K$ spectrum (Fig.~\ref{fig:feeddown_fractions_k_plus}) is significant. The overall feeddown contribution to the $\eta$ spectrum is about 40\% for $p_T < 3~\mathrm{GeV}/c$, and for the $K$ spectrum it is slightly larger. The similar size of the feeddown contribution for $K$ and $\eta$ may help explain why the Ren--Drees method, which does not explicitly include feeddown, gives a prediction for the low-$p_T$ $\eta/\pi^0$ ratio similar to that of our blast-wave fit with feeddown.

The uncertainty band around the fitted $\eta/\pi^0$ parameterization is constructed as follows. For $p_T \lesssim 2.5~\mathrm{GeV}/c$, we use the assigned relative uncertainty of $\pm 5\%$ for the double-ratio method discussed in Sec.~\ref{sec:systematics}. In the intermediate region $2.5 < p_T < 4.0~\mathrm{GeV}/c$, where no pseudo-data points are imposed, the relative uncertainty is interpolated smoothly to the average relative experimental uncertainty of the measured $\eta/\pi^0$ data in the range $3 < p_T < 4~\mathrm{GeV}/c$. For $4 < p_T < 5~\mathrm{GeV}/c$, the band is interpolated linearly to a constant asymptotic uncertainty of 5\%, which is then used for $p_T > 5~\mathrm{GeV}/c$. This construction is intended as a pragmatic uncertainty model that connects the low-$p_T$ model-constrained region smoothly to the experimentally constrained and high-$p_T$ regions.

Figure~\ref{fig:eta_to_pi0_2p76_0_10} also shows a parameterization of the $\eta/\pi^0$ ratio in pp collisions using the parameterization of Eq.~\ref{eq:eta_to_pi0_parameterization}. The parameters used here were obtained from a fit to pp data at different energies while keeping the high-$p_T$ ratio $r = 0.487 \pm 0.024$ \cite{Ren:2021pzi} and the radial flow velocity $\beta = 0$ fixed (see Fig.~\ref{fig:eta_to_pi0_pp_pA_fit}). The difference between the pp parameterization and the Pb--Pb result at low $p_T$ illustrates the significant effect of radial flow on the $\eta/\pi^0$ ratio in heavy-ion collisions.  

\section{Implications for the Decay-Photon Cocktail}

\begin{figure}[t]
\centering
\includegraphics[width=\columnwidth]{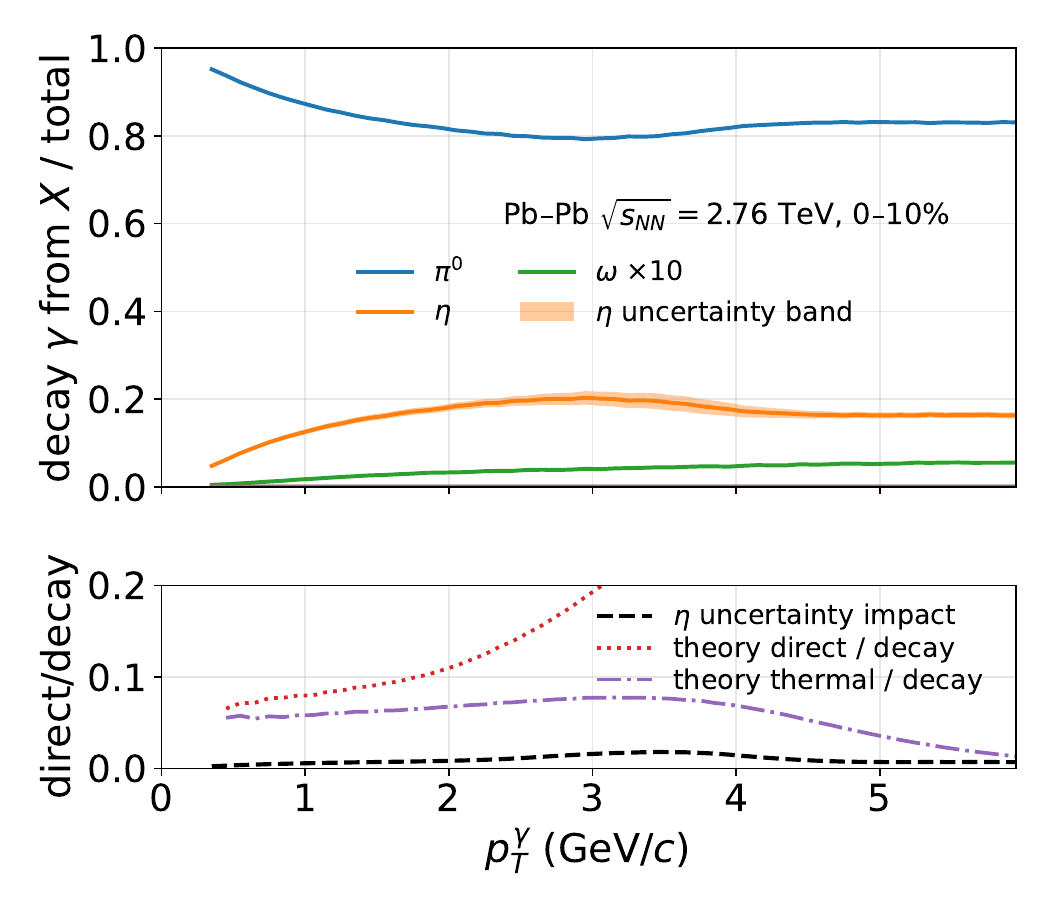}
\caption{Decay-photon cocktail fractions for 0--10\% Pb--Pb collisions at $\sqrt{s_{NN}}=2.76$ TeV. Upper panel: fractional decay-photon contributions from $\pi^0$, $\eta$, and $\omega$. The uncertainty of the $\eta$ contribution is shown as a shaded band. Lower panel: $\eta$-related uncertainty of the decay-photon background compared with the direct-photon signal predicted in Ref.~\cite{Gale:2021emg}.}
\label{fig:decay_photon_cocktail_fractions_2p76_0_10}
\end{figure}

A reduction and improved control of the $\eta/\pi^0$ uncertainty, aimed at enabling more precise direct-photon measurements at low $p_T$, constitutes the primary motivation of this study. The impact of the $\eta/\pi^0$ uncertainty on the direct-photon measurement is illustrated in Figure~\ref{fig:decay_photon_cocktail_fractions_2p76_0_10}. The dominant contributions to the decay-photon background arise from $\pi^0$, $\eta$, and $\omega$ meson decays.

Analogous to the feeddown treatment in the blast-wave + feeddown model, we construct the decay-photon cocktail using precomputed decay-photon kernels. For each hadron with a direct-photon decay branch, the corresponding kernel is generated with PYTHIA~8. The total decay-photon $p_T$ spectrum is then obtained by folding these two-dimensional kernels with the parent-hadron $p_T$ spectra. In this framework, we include contributions from $\pi^0$, $\eta$, $\omega$, $\eta'$, $\rho^0$, $\Sigma^0$, and $\bar{\Sigma}^0$. The $\pi^0$ parameterization is taken from \cite{ALICE:2018mdl}. For hadrons other than $\pi^0$ and $\eta$, we use flow-scaling parameterizations.

Although this method does not represent a substantial conceptual improvement over a standard full Monte Carlo decay simulation, it offers significant practical advantages. In particular, it enables a rapid and flexible evaluation of how variations in the parameterizations of the hadron $p_T$ spectra propagate to the decay-photon cocktail and, consequently, to the extracted direct-photon signal.

The upper panel of Fig.~\ref{fig:decay_photon_cocktail_fractions_2p76_0_10} shows the fractional contributions of selected parent particles to the decay-photon cocktail in 0--10\% Pb--Pb collisions at $\sqrt{s_{NN}}=2.76$ TeV. Clearly, the $\pi^0$ and $\eta$ decay contributions dominate the cocktail across the entire $p_T$ range. On a linear scale, one can see the small contribution of the $\omega$ meson, while all other contributions are negligible. The lower panel illustrates the impact of the $\eta$-related uncertainty on the direct-photon signal, compared with the direct-photon excess predicted by the model calculation of Ref.~\cite{Gale:2021emg}. The total predicted direct-photon signal, comprising prompt, pre-equilibrium, and thermal photons, is shown, along with the thermal-photon contribution separately, for the scenario without suppression from quarks initially being out of chemical equilibrium. Quantitatively, at $p_T \approx 1~\mathrm{GeV}/c$, the $\eta$-related uncertainty corresponds to about 10\% of the expected direct-photon signal. This comparison underscores the importance of constraining the $\eta/\pi^0$ ratio for improving the precision of direct-photon measurements in heavy-ion collisions.

\section{Conclusions}

We have presented a practical, data-driven approach to predict the $\eta/\pi^0$ ratio at low $p_T$ ($p_T \lesssim 3~\mathrm{GeV}/c$), based on the measured charged $K/\pi$ ratio and model input from a blast-wave framework with feeddown contributions. After determining the model parameters from fits to the measured $\pi$, $K$, and $p$ spectra, we construct the $(\eta/\pi^0)/(K^+/\pi^+)$ double ratio and use it to obtain pseudo-data constraints for $\eta/\pi^0$. This method is expected to be useful for future direct-photon and dilepton analyses in heavy-ion collisions, where precise hadron background estimates are essential. We have illustrated this impact for direct photons in Pb--Pb collisions at $\sqrt{s_{NN}}=2.76$~TeV.

The double-ratio method yields predictions that are numerically close to the Ren--Drees approach. This agreement is non-trivial and supports the practical validity of the Ren--Drees method in the studied kinematic range. At the same time, by explicitly modeling the feeddown contributions to the $\eta$ and $K$ spectra and accounting for the particle-mass dependence of radial flow, the present framework makes the underlying ingredients explicit. The flow-scaling cross-check can be viewed as a practical extension of traditional $m_T$ scaling in the presence of strong radial flow and can serve as a more physically motivated baseline for modeling $p_T$ spectra of heavier hadrons in decay-cocktail calculations. In summary, our approach allows for a controlled extrapolation of the $\eta$ meson spectrum toward low $p_T$, addressing a typical bottleneck in direct-photon and dielectron analyses. 

\section*{Data and Code Availability}
No new experimental data were created in this study. The experimental data analyzed in this work were taken from the published sources cited in the text. The analysis code and files used to generate the results and figures are publicly available in Ref.~\cite{reygers_eta_pi0_blastwave_feeddown_2026}. Setup instructions and the analysis workflow are documented in the repository README.

\appendix

\section{\boldmath $\eta/\pi^0$ parameterization in pp collisions}
\begin{figure}[t]
\centering
\includegraphics[width=\columnwidth]{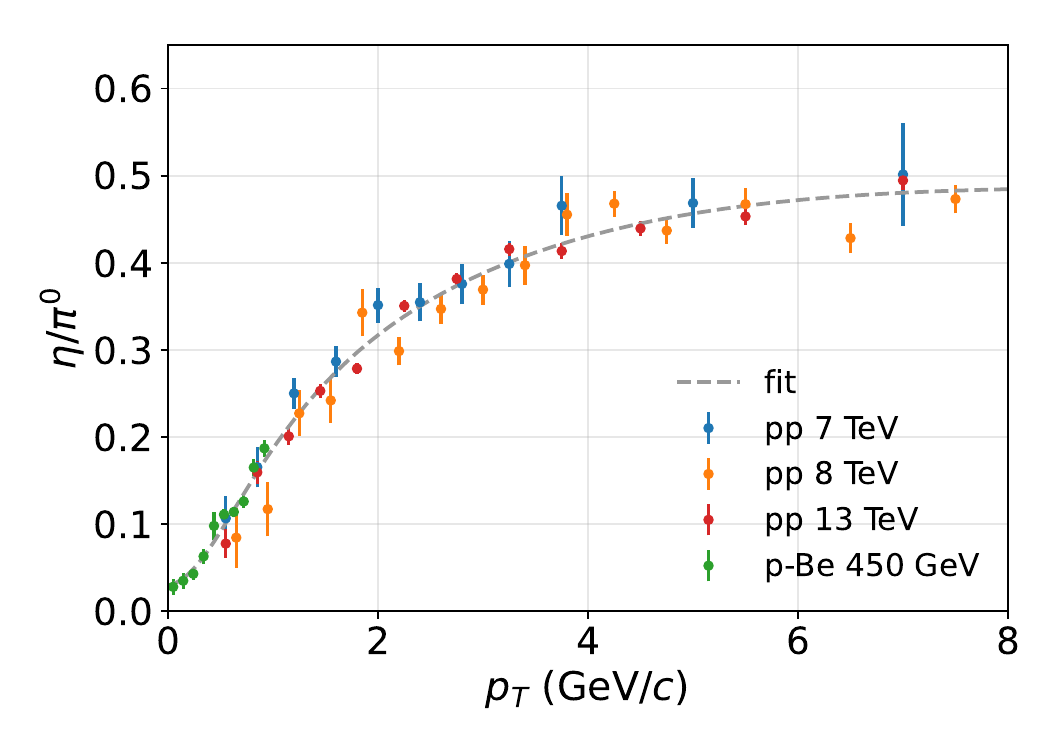}
\caption{Fit of the $\eta/\pi^0$ parameterization to pp and p--A reference data (pp at 7, 8, and 13 TeV, and p--Be at 450 GeV/$c$). The fit uses the same flow-inspired functional form employed in this work for the $\eta/\pi^0$ description. The high-$p_T$ value is fixed to $r=0.487$ and the radial-flow parameter is set to $\beta=0$.}
\label{fig:eta_to_pi0_pp_pA_fit}
\end{figure}

Fig.~\ref{fig:eta_to_pi0_pp_pA_fit} shows how the $\eta/\pi^0$ parameterization for pp collisions used in this paper is obtained from a fit to pp data at $\sqrt{s}=7~\mathrm{TeV}$ \cite{ALICE:2012wos}, $\sqrt{s}=8~\mathrm{TeV}$ \cite{ALICE:2017ryd}, and $\sqrt{s}=13~\mathrm{TeV}$ \cite{ALICE:2024vgi}, as well as fixed-target p--Be data at a beam momentum of 450~GeV/$c$. The $\eta/\pi^0$ ratio as a function of $p_T$ is found to be approximately independent of the center-of-mass energy \cite{Ren:2021pzi}. The data are fitted using Eq.~\ref{eq:eta_to_pi0_parameterization}. The high-$p_T$ $\eta/\pi^0$ value is fixed to $r=0.487$ reported in \cite{Ren:2021pzi}, and the radial-flow velocity is set to $\beta = 0$. The fitted parameters are $A=0.543$, $T=0.107~\mathrm{GeV}$, $p_0=4.60~\mathrm{GeV}/c$, and $d=3.68~\mathrm{GeV}/c$.

\section{\boldmath $\eta/\pi^0$ from $m_T$ scaling}
As an additional baseline, we compare our data-driven $\eta/\pi^0$ result with the traditional $m_T$-scaling approach. In this approach, invariant yields of different hadron species are assumed to have the same functional dependence on transverse mass, up to a constant normalization factor. Using a parameterization $h_{\pi^0}(p_T)$ of the invariant neutral pion $p_T$ spectrum, this corresponds to
\begin{equation}
\left(\frac{\eta}{\pi^0}\right)_{m_T}(p_T)
=
C_{\eta/\pi^0}\,
\frac{h_{\pi^0}\!\left(\sqrt{p_T^2+m_\eta^2-m_{\pi^0}^2}\right)}
     {h_{\pi^0}(p_T)} \, .
\end{equation}
In this work, we fix the normalization to the high-$p_T$ value $C_{\eta/\pi^0}=0.487$.

\begin{figure}[t]
\centering
\includegraphics[width=\columnwidth]{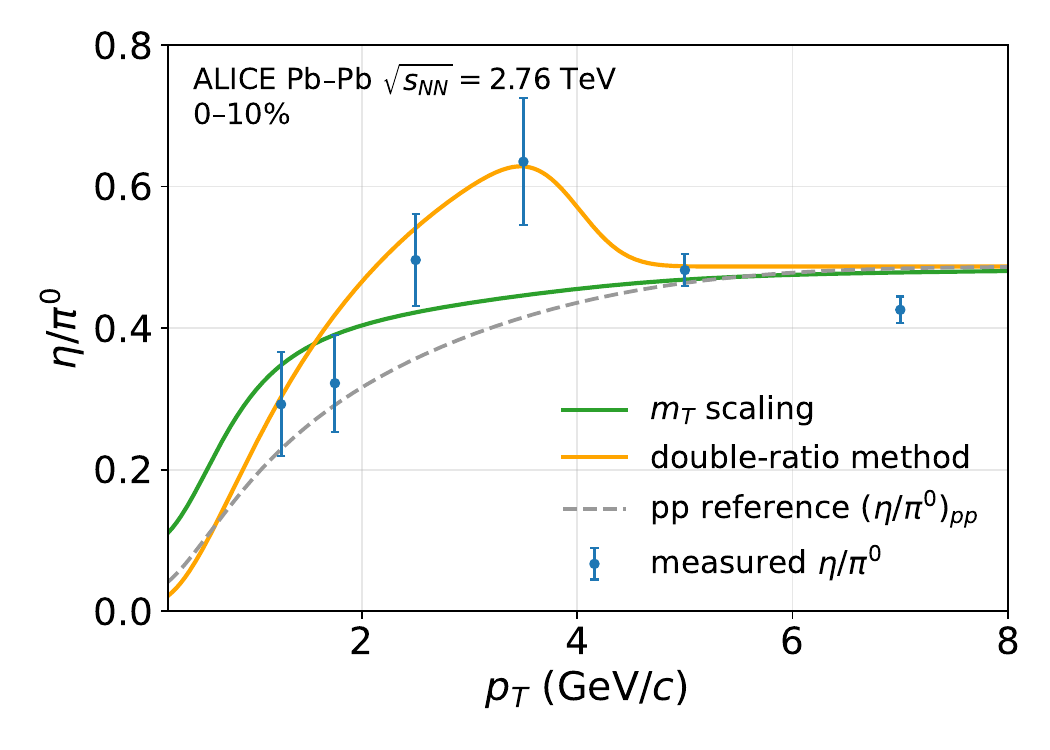}
\caption{Comparison of the $m_T$-scaling prediction (green) and the $\eta/\pi^0$ result from the double-ratio method presented in this work (orange) for 0--10\% Pb--Pb collisions at $\sqrt{s_{NN}}=2.76$~TeV. Measured $\eta/\pi^0$ points are shown in blue; the gray dashed line indicates the pp reference parameterization.}
\label{fig:eta_pi0_mt_vs_flow_2p76_0_10}
\end{figure}

Figure~\ref{fig:eta_pi0_mt_vs_flow_2p76_0_10} compares, for 0--10\% Pb--Pb collisions at $\sqrt{s_{NN}}=2.76$~TeV, the $m_T$-scaling prediction with the measured $\eta/\pi^0$ data and with the result of the double-ratio method presented in this work. The $m_T$-scaling curve is constructed using the $\pi^0$ parameterization from \cite{ALICE:2018mdl}. While $m_T$ scaling provides a useful reference, it does not reproduce the full low-$p_T$ behavior expected in the presence of strong radial flow.

\section{Flow scaling function}

\begin{figure}[t]
\centering
\includegraphics[width=\columnwidth]{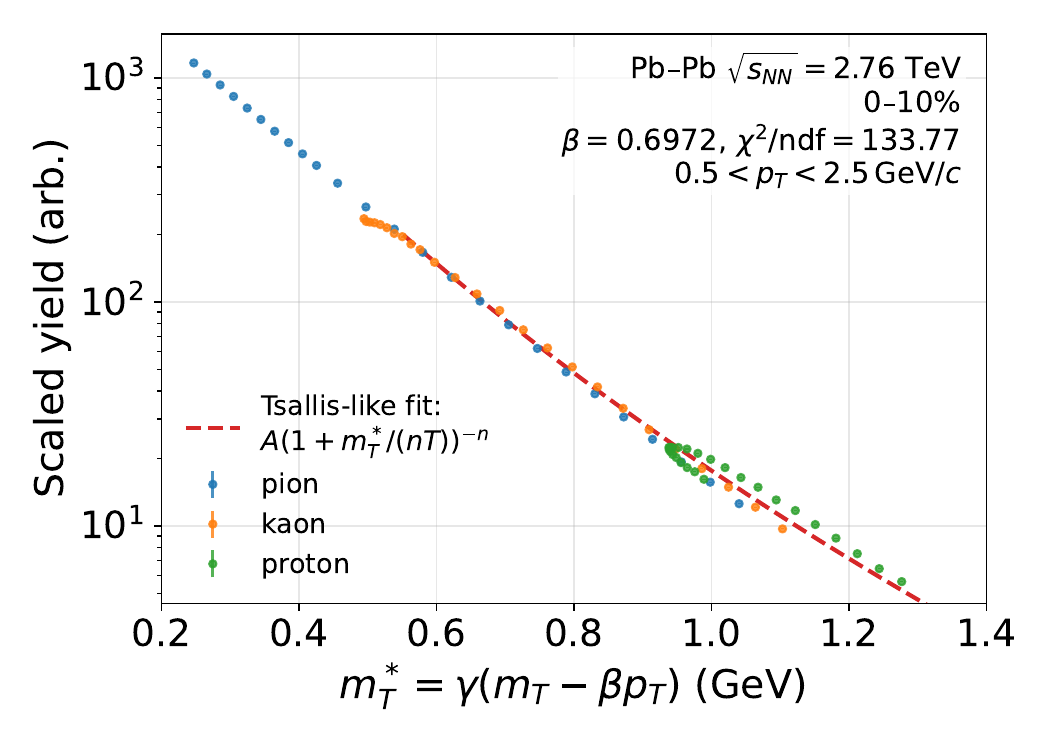}
\caption{Flow-scaling collapse plot for 0--10\% Pb--Pb collisions at $\sqrt{s_{NN}}=2.76$ TeV. The scaled $\pi$, $K$, and $p$ data are shown as a function of $m_T^* = \gamma(m_T-\beta p_T)$, i.e.\ the transverse mass in the source rest system, together with the smooth fit used in the flow-scaling cross-check.}
\label{fig:flow_scaling_collapse_2p76_0_10}
\end{figure}

For 0--10\% central Pb--Pb collisions at $\sqrt{s_{NN}}=2.76$~TeV, the flow-scaling construction gives the best collapse of the measured $\pi$, $K$, and $p$ spectra for an effective radial-flow velocity of $\beta = 0.6972$. The value of $\beta$ is obtained from a scan that minimizes the spread of the scaled hadron spectra in the fit interval $0.5 < p_T < 2.5~\mathrm{GeV}/c$. The corresponding scaling function $F(m_T^*)$ is shown in Fig.~\ref{fig:flow_scaling_collapse_2p76_0_10}, where
\begin{equation}
  m_T^*=\gamma(m_T-\beta p_T)
\end{equation}
denotes the transverse mass in the source rest system. The collapsed yields are parameterized with the Tsallis-like form
\begin{equation}
F(m_T^*) \propto \left(1 + \frac{m_T^*}{n T}\right)^{-n} \,.
\end{equation}
In Fig.~\ref{fig:flow_scaling_collapse_2p76_0_10} we show the range $0.2 \le m_T^* \le 1.4$, which covers the kinematic region relevant for the low-$p_T$ $\eta/\pi^0$ extrapolation. The smooth $F(m_T^*)$ parameterization is used both for the qualitative double-ratio cross-check shown in Fig.~\ref{fig:systematics_summary_2p76_0_10} and for the parent-hadron spectra entering the decay-photon cocktail, as described above. Because the double-ratio cross-check does not implement the same resonance feeddown treatment as the baseline calculation, its deviation is not included in the assigned systematic uncertainty.

\section{\boldmath Feeddown Structure for $\eta$ and $K^+$}
\label{sec:feeddown_structure}

\begin{figure}[t]
\centering
\vspace*{1.5mm}
\includegraphics[width=\columnwidth]{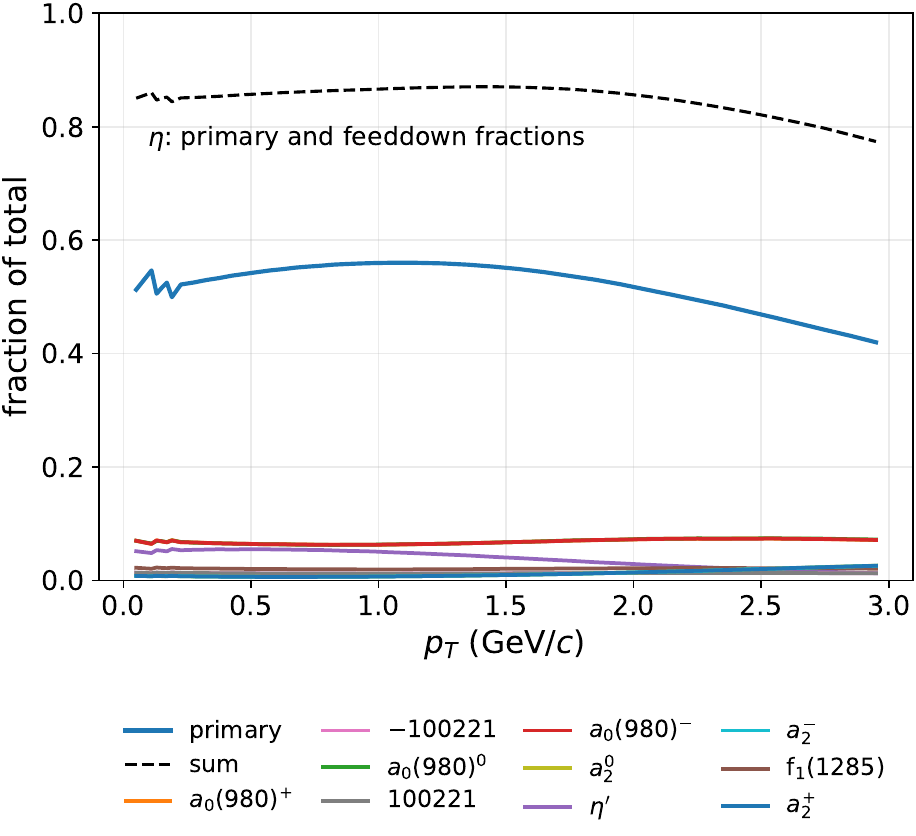}
\caption{Primary and feeddown fractions for $\eta$ as a function of $p_T$ in the blast-wave + feeddown setup used in this work. The ten most dominant parent contributions are shown together with the primary component.}
\label{fig:feeddown_fractions_eta}
\end{figure}

\begin{figure}[t]
\centering
\vspace*{1.5mm}
\includegraphics[width=\columnwidth]{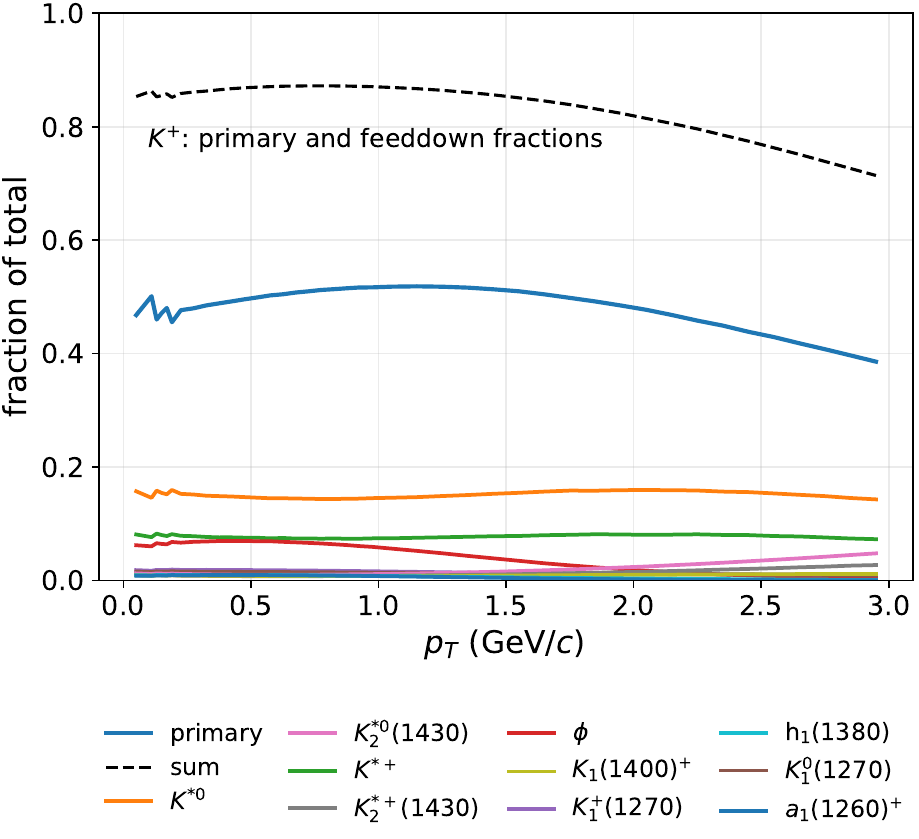}
\caption{Primary and feeddown fractions for $K^+$ as a function of $p_T$ in the blast-wave + feeddown setup used in this work. The ten most dominant parent contributions are shown together with the primary component.}
\label{fig:feeddown_fractions_k_plus}
\end{figure}

The feeddown compositions are shown in Fig.~\ref{fig:feeddown_fractions_eta} for the $\eta$ meson and in Fig.~\ref{fig:feeddown_fractions_k_plus} for the $K^+$ meson. The feeddown is obtained from the PYTHIA 8 particle list for particles up to a mass of $m=2~\mathrm{GeV}/c^2$ and with proper decay length $c\tau < 1~\mathrm{mm}$. The primary-particle spectra are modeled in the blast-wave + feeddown approach with parameters obtained from fits to the $\pi$, $K$, and $p$ spectra in 0--10\% Pb--Pb collisions at $\sqrt{s_{NN}}=2.76$~TeV.

\clearpage

\begin{acknowledgments}
This work has been supported by the DFG (German Research Foundation) --- Project No.\ 273811115 --- SFB 1225 ISOQUANT.
AM is supported by the DFG through the Emmy Noether Programme (Project No.\ 496831614).
AK is supported by the U.S. Department of Energy, Office of Science, Office of Nuclear Physics, grant No. DE-FG02-05ER41367.
We wish to acknowledge helpful discussions with Ana Marin, Ilya Fokin, and Johanna Stachel. We used OpenAI Codex to assist with code generation based on human-written prompts. All scientific results, validations, and interpretations were performed by the authors.

\end{acknowledgments}

\bibliography{master}

@PREAMBLE{
 "\providecommand{\noopsort}[1]{}" 
 # "\providecommand{\singleletter}[1]{#1}%" 
}

@article{WA98:2000vxl,
    author = "Aggarwal, M. M. and others",
    collaboration = "WA98",
    title = "{Observation of direct photons in central 158-A-GeV Pb-208 + Pb-208 collisions}",
    eprint = "nucl-ex/0006008",
    archivePrefix = "arXiv",
    doi = "10.1103/PhysRevLett.85.3595",
    journal = "Phys. Rev. Lett.",
    volume = "85",
    pages = "3595--3599",
    year = "2000"
}

@article{ALICE:2015xmh,
    author = "Adam, Jaroslav and others",
    collaboration = "ALICE",
    title = "{Direct photon production in Pb-Pb collisions at $\sqrt{s_{NN}} =$ 2.76 TeV}",
    eprint = "1509.07324",
    archivePrefix = "arXiv",
    primaryClass = "nucl-ex",
    reportNumber = "ALICE-PUBLIC-2015-007, CERN-PH-EP-2015-254",
    doi = "10.1016/j.physletb.2016.01.020",
    journal = "Phys. Lett. B",
    volume = "754",
    pages = "235--248",
    year = "2016"
}

@article{STAR:2016use,
    author = "Adamczyk, L. and others",
    collaboration = "STAR",
    title = "{Direct virtual photon production in Au+Au collisions at $\sqrt{s_{NN}}$ = 200 GeV}",
    eprint = "1607.01447",
    archivePrefix = "arXiv",
    primaryClass = "nucl-ex",
    doi = "10.1016/j.physletb.2017.04.050",
    journal = "Phys. Lett. B",
    volume = "770",
    pages = "451--458",
    year = "2017"
}

@article{PHENIX:2008uif,
    author = "Adare, A. and others",
    collaboration = "PHENIX",
    title = "{Enhanced production of direct photons in Au+Au collisions at $\sqrt{s_{NN}}=200$ GeV and implications for the initial temperature}",
    eprint = "0804.4168",
    archivePrefix = "arXiv",
    primaryClass = "nucl-ex",
    doi = "10.1103/PhysRevLett.104.132301",
    journal = "Phys. Rev. Lett.",
    volume = "104",
    pages = "132301",
    year = "2010"
}

@article{Ren:2021pzi,
    author = "Ren, Yuanjie and Drees, Axel",
    title = "{Study of the $\eta$ to $\pi^0$ Ratio in Heavy-Ion Collisions}",
    eprint = "2102.05220",
    archivePrefix = "arXiv",
    primaryClass = "nucl-ex",
    doi = "10.1103/PhysRevC.104.054902",
    journal = "Phys. Rev. C",
    volume = "104",
    number = "5",
    pages = "054902",
    year = "2021"
}

@article{PHENIX:2022rsx,
    author = "Acharya, U. A. and others",
    collaboration = "PHENIX",
    title = "{Nonprompt direct-photon production in Au$+$Au collisions at $\sqrt{s_{_{NN}}}=200$ GeV}",
    eprint = "2203.17187",
    journal = "",
    archivePrefix = "arXiv",
    primaryClass = "nucl-ex",
    month = "3",
    year = "2022"
}

@article{Floerchinger:2018pje,
    author = "Floerchinger, Stefan and Grossi, Eduardo and Lion, Jorrit",
    title = "{Fluid dynamics of heavy ion collisions with mode expansion}",
    eprint = "1811.01870",
    archivePrefix = "arXiv",
    primaryClass = "nucl-th",
    doi = "10.1103/PhysRevC.100.014905",
    journal = "Phys. Rev. C",
    volume = "100",
    number = "1",
    pages = "014905",
    year = "2019"
}

@article{Mazeliauskas:2018irt,
    author = "Mazeliauskas, Aleksas and Floerchinger, Stefan and Grossi, Eduardo and Teaney, Derek",
    title = "{Fast resonance decays in nuclear collisions}",
    eprint = "1809.11049",
    archivePrefix = "arXiv",
    primaryClass = "nucl-th",
    doi = "10.1140/epjc/s10052-019-6791-7",
    journal = "Eur. Phys. J. C",
    volume = "79",
    number = "3",
    pages = "284",
    year = "2019"
}

@article{Vermunt:2023zsk,
    author = "Vermunt, L. and Seemann, Y. and Dubla, A. and Floerchinger, S. and Grossi, E. and Kirchner, A. and Masciocchi, S. and Selyuzhenkov, I.",
    title = "{Mapping properties of the quark gluon plasma in Pb-Pb and Xe-Xe collisions at energies available at the CERN Large Hadron Collider}",
    eprint = "2308.16722",
    archivePrefix = "arXiv",
    primaryClass = "hep-ph",
    doi = "10.1103/PhysRevC.108.064908",
    journal = "Phys. Rev. C",
    volume = "108",
    number = "6",
    pages = "064908",
    year = "2023"
}

@article{ALICE:2014juv,
    author = "Abelev, Betty Bezverkhny and others",
    collaboration = "ALICE",
    title = "{Production of charged pions, kaons and protons at large transverse momenta in pp and Pb{\textendash}Pb collisions at $\sqrt{s_{\rm NN}}$ =2.76 TeV}",
    eprint = "1401.1250",
    archivePrefix = "arXiv",
    primaryClass = "nucl-ex",
    reportNumber = "CERN-PH-EP-2013-230",
    doi = "10.1016/j.physletb.2014.07.011",
    journal = "Phys. Lett. B",
    volume = "736",
    pages = "196--207",
    year = "2014"
}

@article{ALICE:2018mdl,
    author = "Acharya, Shreyasi and others",
    collaboration = "ALICE",
    title = "{Neutral pion and $\eta$ meson production at mid-rapidity in Pb-Pb collisions at $\sqrt{s_{NN}}$ = 2.76 TeV}",
    eprint = "1803.05490",
    archivePrefix = "arXiv",
    primaryClass = "nucl-ex",
    reportNumber = "CERN-EP-2018-040",
    doi = "10.1103/PhysRevC.98.044901",
    journal = "Phys. Rev. C",
    volume = "98",
    number = "4",
    pages = "044901",
    year = "2018"
}

@article{ALICE:2024vgi,
    author = "Acharya, Shreyasi and others",
    collaboration = "ALICE",
    title = "{Light neutral-meson production in pp collisions at $\sqrt{s}$ = 13 TeV}",
    eprint = "2411.09560",
    archivePrefix = "arXiv",
    primaryClass = "hep-ex",
    reportNumber = "CERN-EP-2024-304",
    doi = "10.1007/JHEP08(2025)035",
    journal = "JHEP",
    volume = "08",
    pages = "035",
    year = "2025"
}

@article{Sjostrand:2014zea,
    author = "Sjostrand, Torbjorn and Ask, Stefan and Christiansen, Jesper R. and Corke, Richard and Desai, Nishita and Ilten, Philip and Mrenna, Stephen and Prestel, Stefan and Rasmussen, Christine O. and Skands, Peter Z.",
    title = "{An Introduction to PYTHIA 8.2}",
    eprint = "1410.3012",
    archivePrefix = "arXiv",
    primaryClass = "hep-ph",
    reportNumber = "LU-TP-14-20, MCNET-14-11, CERN-PH-TH-2014-190",
    doi = "10.1016/j.cpc.2015.01.024",
    journal = "Comput. Phys. Commun.",
    volume = "191",
    pages = "159--177",
    year = "2015"
}

@article{Albrecht:1991zza,
    author = "Albrecht, R. and others",
    collaboration = "WA80",
    title = "{Production of $\pi^0$, $\eta$ and direct photons in O+Au collisions at 200A GeV}",
    journal = "Phys. Rev. C",
    volume = "44",
    pages = "2736--2752",
    year = "1991",
    doi = "10.1103/PhysRevC.44.2736"
}

@article{Cleymans:1999st,
    author = "Cleymans, J. and Redlich, K.",
    title = "{Unified description of freezeout parameters in relativistic heavy ion collisions}",
    eprint = "nucl-th/9903063",
    archivePrefix = "arXiv",
    reportNumber = "KUL-TF-99-06",
    doi = "10.1103/PhysRevC.60.054908",
    journal = "Phys. Rev. C",
    volume = "60",
    pages = "054908",
    year = "1999"
}

@article{Andronic:2017pug,
    author = "Andronic, A. and Braun-Munzinger, P. and Redlich, K. and Stachel, J. and others",
    title = "{Decoding the phase structure of QCD via particle production at high energy}",
    journal = "Nature",
    volume = "561",
    pages = "321--330",
    year = "2018",
    doi = "10.1038/s41586-018-0491-6",
    eprint = "1710.09425",
    archivePrefix = "arXiv",
    primaryClass = "nucl-th"
}

@article{David:2019wpt,
    author = "David, Gabor",
    title = "{Direct real photons in relativistic heavy ion collisions}",
    eprint = "1907.08893",
    archivePrefix = "arXiv",
    primaryClass = "nucl-ex",
    doi = "10.1088/1361-6633/ab6f57",
    journal = "Rept. Prog. Phys.",
    volume = "83",
    number = "4",
    pages = "046301",
    year = "2020"
}

@article{ALICE:2012wos,
    author = "Abelev, B. and others",
    collaboration = "ALICE",
    title = "{Neutral pion and $\eta$ meson production in proton-proton collisions at $\sqrt{s}=0.9$ TeV and $\sqrt{s}=7$ TeV}",
    eprint = "1205.5724",
    archivePrefix = "arXiv",
    primaryClass = "hep-ex",
    reportNumber = "CERN-PH-EP-2012-001",
    doi = "10.1016/j.physletb.2012.09.015",
    journal = "Phys. Lett. B",
    volume = "717",
    pages = "162--172",
    year = "2012"
}

@article{ALICE:2017ryd,
    author = "Acharya, Shreyasi and others",
    collaboration = "ALICE",
    title = "{$\pi ^{0}$ and $\eta $ meson production in proton-proton collisions at $\sqrt{s}=8$ TeV}",
    eprint = "1708.08745",
    archivePrefix = "arXiv",
    primaryClass = "hep-ex",
    reportNumber = "CERN-EP-2017-216",
    doi = "10.1140/epjc/s10052-018-5612-8",
    journal = "Eur. Phys. J. C",
    volume = "78",
    number = "3",
    pages = "263",
    year = "2018"
}

@article{Schnedermann:1992ra,
      author         = "Schnedermann, Ekkard and Sollfrank, Josef and Heinz,
                        Ulrich W.",
      title          = "{Fireball spectra}",
      booktitle      = "{NATO Advanced Study Institute on Particle Production in
                        Highly Excited Matter Castelvecchio Pascoli, Italy, July
                        12-24, 1992}",
      journal        = "NATO Sci. Ser. B",
      volume         = "303",
      year           = "1993",
      pages          = "175-206",
      reportNumber   = "TPR-92-29",
      SLACcitation   = "%%CITATION = TPR-92-29;%%"
}

@article{Schnedermann:1993ws,
      author         = "Schnedermann, Ekkard and Sollfrank, Josef and Heinz,
                        Ulrich W.",
      title          = "{Thermal phenomenology of hadrons from 200-A/GeV S+S
                        collisions}",
      journal        = "Phys. Rev.",
      volume         = "C48",
      year           = "1993",
      pages          = "2462-2475",
      doi            = "10.1103/PhysRevC.48.2462",
      eprint         = "nucl-th/9307020",
      archivePrefix  = "arXiv",
      primaryClass   = "nucl-th",
      reportNumber   = "TPR-93-16",
      SLACcitation   = "%%CITATION = NUCL-TH/9307020;%%"
}

@article{Cooper:1974mv,
      author         = "Cooper, Fred and Frye, Graham",
      title          = "{Comment on the Single Particle Distribution in the
                        Hydrodynamic and Statistical Thermodynamic Models of
                        Multiparticle Production}",
      journal        = "Phys. Rev.",
      volume         = "D10",
      year           = "1974",
      pages          = "186",
      doi            = "10.1103/PhysRevD.10.186",
      reportNumber   = "Print-74-0742 (YESHIVA)",
      SLACcitation   = "%%CITATION = PHRVA,D10,186;%%"
}

@article{Mazeliauskas:2019ifr,
    author = "Mazeliauskas, Aleksas and Vislavicius, Vytautas",
    title = "{Temperature and fluid velocity on the freeze-out surface from $\pi$, $K$, $p$ spectra in pp, p-Pb and Pb-Pb collisions}",
    eprint = "1907.11059",
    archivePrefix = "arXiv",
    primaryClass = "hep-ph",
    doi = "10.1103/PhysRevC.101.014910",
    journal = "Phys. Rev. C",
    volume = "101",
    number = "1",
    pages = "014910",
    year = "2020"
}

@article{Lu:2024shm,
    author = "Lu, Pengzhong and Kavak, Rafet and Dubla, Andrea and Masciocchi, Silvia and Selyuzhenkov, Ilya",
    title = "{Quantification of the low-$p_\textrm{T}$ pion excess in heavy-ion collisions at the LHC and top RHIC energy}",
    eprint = "2407.09207",
    archivePrefix = "arXiv",
    primaryClass = "hep-ph",
    doi = "10.1007/s41365-025-01718-z",
    journal = "Nucl. Sci. Tech.",
    volume = "36",
    number = "8",
    pages = "142",
    year = "2025"
}

@article{Nijs:2020roc,
    author = {Nijs, Govert and van der Schee, Wilke and G{\"u}rsoy, Umut and Snellings, Raimond},
    title = "{Bayesian analysis of heavy ion collisions with the heavy ion computational framework Trajectum}",
    eprint = "2010.15134",
    archivePrefix = "arXiv",
    primaryClass = "nucl-th",
    reportNumber = "CERN-TH-2020-175, MIT-CTP/5251",
    doi = "10.1103/PhysRevC.103.054909",
    journal = "Phys. Rev. C",
    volume = "103",
    number = "5",
    pages = "054909",
    year = "2021"
}

@article{Busza:2018rrf,
    author = "Busza, Wit and Rajagopal, Krishna and van der Schee, Wilke",
    title = "{Heavy Ion Collisions: The Big Picture, and the Big Questions}",
    eprint = "1802.04801",
    archivePrefix = "arXiv",
    primaryClass = "hep-ph",
    reportNumber = "MIT-CTP-4892, MIT-CTP/4892",
    doi = "10.1146/annurev-nucl-101917-020852",
    journal = "Ann. Rev. Nucl. Part. Sci.",
    volume = "68",
    pages = "339--376",
    year = "2018"
}

@article{Gale:2021emg,
    author = {Gale, Charles and Paquet, Jean-Fran{\c{c}}ois and Schenke, Bj{\"o}rn and Shen, Chun},
    title = "{Multimessenger heavy-ion collision physics}",
    eprint = "2106.11216",
    archivePrefix = "arXiv",
    primaryClass = "nucl-th",
    doi = "10.1103/PhysRevC.105.014909",
    journal = "Phys. Rev. C",
    volume = "105",
    number = "1",
    pages = "014909",
    year = "2022"
}

@article{Vovchenko:2019pjl,
    author = "Vovchenko, Volodymyr and Stoecker, Horst",
    title = "{Thermal-FIST: A package for heavy-ion collisions and hadronic equation of state}",
    eprint = "1901.05249",
    archivePrefix = "arXiv",
    primaryClass = "nucl-th",
    doi = "10.1016/j.cpc.2019.06.024",
    journal = "Comput. Phys. Commun.",
    volume = "244",
    pages = "295--310",
    year = "2019"
}

@misc{reygers_eta_pi0_blastwave_feeddown_2026,
    author = "Kirchner, Andreas and Mazeliauskas, Aleksas and Reygers, Klaus",
    title = "{eta-pi0-blastwave-feeddown}",
    year = "2026",
    howpublished = "\url{https://github.com/reygers/eta-pi0-blastwave-feeddown}",
    note = "GitHub repository, accessed 2026-03-13"
}

@article{Borsanyi:2018grb,
    author = "Borsanyi, Szabolcs and Fodor, Zoltan and Guenther, Jana N. and Katz, Sandor K. and Szabo, Kalman K. and Pasztor, Attila and Portillo, Israel and Ratti, Claudia",
    title = "{Higher order fluctuations and correlations of conserved charges from lattice QCD}",
    eprint = "1805.04445",
    archivePrefix = "arXiv",
    primaryClass = "hep-lat",
    doi = "10.1007/JHEP10(2018)205",
    journal = "JHEP",
    volume = "10",
    pages = "205",
    year = "2018"
}

@article{HotQCD:2018pds,
    author = "Bazavov, A. and others",
    collaboration = "HotQCD",
    title = "{Chiral crossover in QCD at zero and non-zero chemical potentials}",
    eprint = "1812.08235",
    archivePrefix = "arXiv",
    primaryClass = "hep-lat",
    doi = "10.1016/j.physletb.2019.05.013",
    journal = "Phys. Lett. B",
    volume = "795",
    pages = "15--21",
    year = "2019"
}

@software{FastReso,
    author = "Mazeliauskas, Aleksas and Floerchinger, Stefan and Grossi, Eduardo and Teaney, Derek",
    title = "{FastReso}",
    year = "2018",
    url = "https://github.com/amazeliauskas/FastReso",
    license = "MIT"
}

\end{document}